\documentclass[prl,reprint,twocolumn,superscriptaddress,floatfix]{revtex4-2}

\usepackage{amsmath}
\usepackage{amsfonts}
\usepackage{amssymb}
\usepackage{mathtools}
\usepackage{empheq}
\usepackage{multirow}
\usepackage{lipsum}
\usepackage{indentfirst}
\usepackage{authoraftertitle}
\usepackage{titlesec}

\usepackage{graphicx}
\usepackage[usenames,dvipsnames]{xcolor}

\usepackage[varg]{txfonts}
\usepackage{newtxtext}

\usepackage{bm}

\usepackage{physics}
\let\exp\exponential
\let\ln\naturallogarithm

\definecolor{pdarkblue}{rgb}{0.1797, 0.1875, 0.5703}
\usepackage{hyperref}
\hypersetup{
    colorlinks=true,
    citecolor=pdarkblue,
    linkcolor=pdarkblue,
    urlcolor=pdarkblue,
}
\usepackage[all]{hypcap}

\DeclareMathAlphabet{\mathsfit}{T1}{\sfdefault}{\mddefault}{\sldefault}

\DeclareMathOperator*{\artanh}{artanh}
\let\Re\relax
\let\Im\relax
\DeclareMathOperator*{\Re}{Re}
\DeclareMathOperator*{\Im}{Im}

\newcommand\R{\mathsfit{R}}
\newcommand\imag{{\mathrm{i}}}
\newcommand\cur{\mathcal{J}}
\newcommand\act{\mathcal{A}}
\newcommand\aff{\mathcal{F}}
\newcommand\flux{\mathcal{T}}
\newcommand\st{q}
\newcommand\lr{\lambda^{\mathrm{R}}}
\newcommand\li{\lambda^{\mathrm{I}}}
\newcommand{\ave}[1]{\langle #1 \rangle}
\newcommand{\qave}[1]{\left\langle #1 \right\rangle}

\newcommand{\kBphys}{k_{\mathrm B}}
\newcommand{\largestar}{\mathchoice{\textstyle{*}}{\textstyle{*}}{\textstyle{*}}{\scriptstyle{*}}}
\newcommand{\CS}{{\mathcal C^{\largestar}}}
\newcommand{\Cuni}{{\mathcal C_\mathrm{uni}}}
\newcommand{\Casy}{{\mathcal C_\mathrm{asy}}}
\newcommand{\Eplus}{{\mathcal E^+}}
\newcommand{\qed}{\hfill$\blacksquare$\vspace{\sectionsep}}
\newcommand{\revecA}{\bm{u}^{(\alpha)}}
\newcommand{\levecA}{\bm{v}^{(\alpha)}}
\newcommand{\asymm}{\chi_{ba}}
\newcommand{\Bvert}[1]{\Biggl\lvert #1 \Biggr\rvert}
\newcommand{\Bparen}[1]{\Biggl( #1 \Biggr)}

\newlength{\sectionsep}
\renewcommand{\paragraph}[1]{\vspace{\sectionsep}\textit{#1}.---\ignorespaces}
\allowdisplaybreaks

\def\SetMake#1|#2?{\left\{#1\,\middle|\,#2\right\}}
\newcommand{\Set}[1]{{\SetMake#1?}}

\begin{document}

\title{Thermodynamic Bound on the Asymmetry of Cross-Correlations}
\date{\today}

\author{Naruo Ohga}
\email{naruo.ohga@ubi.s.u-tokyo.ac.jp}
\affiliation{Department of Physics, Graduate School of Science, The University of Tokyo, 7-3-1 Hongo, Bunkyo-ku, Tokyo 113-0033, Japan}

\author{Sosuke Ito}
\affiliation{Department of Physics, Graduate School of Science, The University of Tokyo, 7-3-1 Hongo, Bunkyo-ku, Tokyo 113-0033, Japan}
\affiliation{Universal Biology Institute, Graduate School of Science, The University of Tokyo, 7-3-1 Hongo, Bunkyo-ku, Tokyo 113-0033, Japan}

\author{Artemy Kolchinsky}
\affiliation{Universal Biology Institute, Graduate School of Science, The University of Tokyo, 7-3-1 Hongo, Bunkyo-ku, Tokyo 113-0033, Japan}

\begin{abstract}
The principle of microscopic reversibility says that, in equilibrium, two-time cross-correlations are symmetric under the exchange of observables. Thus, the asymmetry of cross-correlations is a fundamental, measurable, and often-used statistical signature of deviation from equilibrium. Here we find a simple and universal inequality that bounds the magnitude of asymmetry by the cycle affinity, i.e., the strength of thermodynamic driving. Our result applies to a large class of systems and all state observables, and it suggests a fundamental thermodynamic cost for various nonequilibrium functions quantified by the asymmetry. It also provides a powerful tool to infer affinity from measured cross-correlations, in a different and complementary way to the thermodynamic uncertainty relations. As an application, we prove a thermodynamic bound on the coherence of noisy oscillations, which was previously conjectured by Barato and Seifert [\href{https://doi.org/10.1103/PhysRevE.95.062409}{Phys.~Rev.~E \textbf{95}, 062409 (2017)}]. We also derive a thermodynamic bound on directed information flow in a biochemical signal transduction model.
\end{abstract}

\maketitle

\fontsize{10.5pt}{12.36pt}\selectfont
\fontdimen2\font=2.9pt 

\begin{figure}[b]
    \centering
    \includegraphics[width=\hsize,clip]{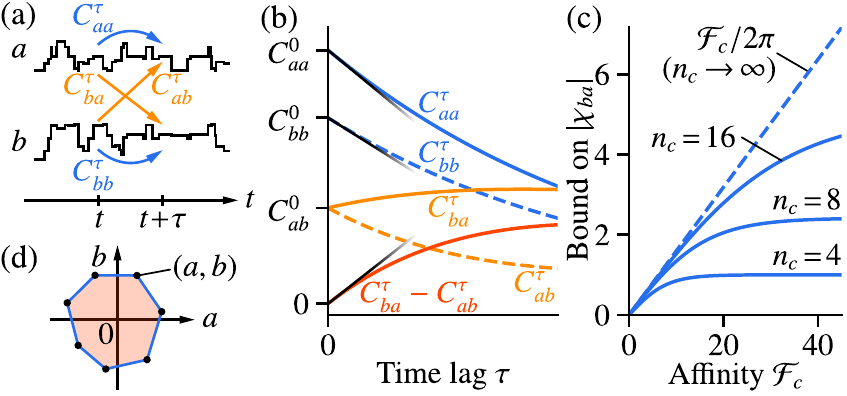}
    \caption{(a) A pair of observables $a(t)$ and $b(t)$ is measured in a stochastic system in steady state, from which the two-time correlations are calculated.
    (b) We define a normalized measure of asymmetry of cross-correlation $\asymm$ in Eq.~\eqref{eq:asymm_def} based on the short-time behavior of the correlation functions (black).
    (c) Our main result is a thermodynamic relation between $\asymm$ and cycle affinity, Eq.~\eqref{eq:main}. 
    (d) The result is derived in part by considering $\asymm$ as the ratio between the area and perimeter of the polygon formed by the values $(a,b)$ over a cycle and then using the isoperimetric inequality.  
    }
\label{fig_illustration}
\end{figure}

One of the most common ways to characterize a physical system is by studying its spatiotemporal correlations.  Imagine measuring a pair of observables $a(t)$ and $b(t)$ in a stochastic system in steady state, e.g., 2 degrees of freedom, counts of two chemical species, fluorescence intensity of two colors, voltages of two points, etc.~[Fig.~\ref{fig_illustration}(a)]. The two-time correlation between $a$ and $b$ at time lag $\tau$ is 
\begin{equation}
    C_{ba}^{\tau} \coloneqq { \qave{ b(t+\tau) a(t) } } ,
\end{equation}
where $\ave{\cdot}$ indicates average across time or trials (it does not depend on $t$ due to the steady-state assumption).  
When $a=b$, $C_{aa}^\tau$  is the autocorrelation function of   $a$.
When $a\ne b$, $C_{ba}^\tau$ captures the cross-correlation from $a$ to $b$.

A classic result in statistical physics states that  cross-correlations  reflect thermodynamic properties of the steady state~\cite{CasimirOnOnsagers1945}. 
In systems without odd degrees of freedom,  cross-correlations in equilibrium  obey the symmetry $ C_{ba}^{\tau}= C_{ab}^{\tau}$ for all $a$, $b$, and $\tau$. This is called the principle of microscopic reversibility, and it serves as the basis of the celebrated Onsager reciprocity~\cite{OnsagerReciprocalI1931,CasimirOnOnsagers1945}.
Thus, violation of this symmetry is a fundamental and often-used statistical signature of nonequilibrium steady states~\cite{steinberg1986time,rothberg2001testing,EckhardtNoiseCorrelations2003,QianFluorescenceCorrelation2004,JachensAsymmetryOfTemporalCrossCorrelations2006,PaneniTemporalAsymmetry2008,choi2008testing,crisanti2012nonequilibrium,battleBrokenDetailedBalance2016,GladrowNonequilibriumDynamics2017,ghantaFluctuationLoops2017,muraNonequilibriumScaling2018,BrogioliAsymmetricTimeCrossCorrelation2019}.

To maintain a nonequilibrium steady state, a system must be subjected to thermodynamic driving, such as a temperature gradient, a chemical potential gradient, or an external force. In discrete-state stochastic systems as considered here, the strength of different kinds of driving can be quantified in a unified way by the ``cycle affinity''~\cite{schnakenberg1976network,seifert2012stochastic}. 
Given a cycle (cyclic sequence of states) $c$, the cycle affinity $\aff_c$ is the sum of the thermodynamic forces acting on the system around the cycle. Equivalently, each time the system completes the cycle, the thermodynamic entropy of the environment increases by $\aff_c$~\footnote{We treat entropy as dimensionless by dividing it by the Boltzmann constant $\kBphys$.}. 
Cycle affinity vanishes in equilibrium, and it is a fundamental thermodynamic signature of nonequilibrium steady states.  Until now, however, these thermodynamic and statistical signatures of deviation from equilibrium have not been universally and quantitatively related.

In this Letter, we prove a universal and simple relationship between asymmetry of cross-correlations and cycle affinity in steady state.  To introduce our result, we first define a normalized measure of asymmetry between $a$ and $b$,
\begin{equation}
    \asymm := \lim_{\tau\to0}
    \frac{ C_{ba}^{\tau}-C_{ab}^{\tau}}{2\sqrt{ (\Delta_\tau C_{aa})(\Delta_\tau C_{bb})}} \,.
    \label{eq:asymm_def}
\end{equation}
This measure is dimensionless and invariant under shifting and scaling of observables and time. 
The normalization terms $ \Delta_\tau C_{aa} \coloneqq C_{aa}^0 - C_{aa}^\tau$ and $ \Delta_\tau C_{bb} \coloneqq C_{bb}^0 - C_{bb}^\tau$  quantify the decay of autocorrelations of $a$ and $b$ [Fig.~\ref{fig_illustration}(b)]. They may also be interpreted as measures of diffusion (mean-squared displacements), since $\Delta_\tau C_{aa} = \frac{1}{2}\ave{ [a(t+\tau) - a(t)]^2 }$.

Our main result is an inequality between cycle affinity and normalized asymmetry between any pair of observables,
\begin{equation}
    \vert \asymm \vert  
    \leq  \max_{c} \frac{ \tanh(\aff_{c}/2n_{c})}{\tan(\pi/n_{c})}
    \le \max_c \frac{\aff_c}{2\pi} \,,
    \label{eq:main}
\end{equation}
where $\max_c$ is the maximum over all simple cycles and $n_c$ is the number of states in cycle $c$ [see Fig.~\ref{fig_illustration}(c)]. The second bound, which refers to cycle affinity per radian, is approached in the limit $n_c\to \infty$ at a fixed cycle affinity. Some tighter versions of Eq.~\eqref{eq:main} are provided below.

Our result is physically meaningful and experimentally accessible. It provides a fundamental bound on the asymmetry of cross-correlations achievable at a given level of affinity. This suggests the existence of thermodynamic trade-offs for various physical functions that can be quantified by such asymmetry,
including directed interactions and information flow~\cite{NolteIdentifyingInteractions2006,SisanEventOrdering2010,BorysovCrossCorrelationAsymmetries2014,kullmann2002time,HaufeCriticalAssessmentOfConnectivity2013,QinBiochemicalNetworkControlling2018}, nonequilibrium oscillations and  circulation~\cite{tomitaIrreversibleCirculation1974,QianConcentrationFluctuations2002,ghantaFluctuationLoops2017,GnesottoBrokenDetailedBalanceReview2018,gonzalezExperimentalMetrics2019,yasudaNonreciprocality2021,teitsworthStochasticLineIntegrals2022}, nonreciprocal motion~\cite{golestanian2008analytic,golestanian2009stochastic,tarama2018mechanics}, and anomalous response (such as odd viscosity)~\cite{evansSymmetryAnalysis1989,EpsteinTimeReversal2020,HargusTimeReversal2020,YasudaTimeCorrelation2022}.

The normalized asymmetry $\asymm$ is also experimentally accessible, since it depends only on the short-time two-time correlation functions. In fact, $\asymm$ can be expressed in terms of the slopes of auto- and cross-correlation functions, $\asymm= (\partial_\tau C_{ba}^{\tau}-\partial_\tau C_{ab}^{\tau})/\bigl[ 2 \!\sqrt{ (\partial_\tau C_{aa}^\tau )( \partial_\tau C_{bb}^\tau)} \bigr]$, where the derivative $\partial_\tau \equiv\partial/\partial \tau$ is evaluated at $\tau = 0$ [Fig.~\ref{fig_illustration}(b)]. 
Such correlations can be measured using, for example, fluorescence cross-correlation spectroscopy~\cite{EigenSortingSingleMolecules1994,SchwilleDualColorFluorescence1997,elson_fluorescence_2011}, microscopy~\cite{battleBrokenDetailedBalance2016,roldan2021quantifying}, and other experimental techniques~\cite{ishii_two-dimensional_2013,talele_reaction_2022}.  In fact, even a single time series can be used if one of the observables is taken to be a nonlinear transformation of the other, such as $b(t)=a(t)^2$. 
Equation~\eqref{eq:main} therefore provides a powerful tool for inferring thermodynamic properties from experimental data, in complement to existing techniques like thermodynamic uncertainty relations (TURs)~\cite{Barato2015thermodynamic,GingrichInferringDissipation2017,SeifertFromStochasticThermodynamics2019,horowitz2020thermodynamic}. We contrast our bound with TURs below.

The derivation of our result, found at the end of this Letter, combines existing techniques from stochastic thermodynamics with some new ideas. Our most important new idea is to interpret Eq.~\eqref{eq:asymm_def} as the ratio between the area and circumference of the polygon swept out by (appropriately scaled) observables $a$ and $b$ over a cycle [Fig.~\ref{fig_illustration}(d)]. We then employ the isoperimetric inequality~\cite{OssermanIsoperimetricInequality1978}, which says that the area of $n$-sided polygon with a given circumference is maximized by the regular $n$-sided polygon~\footnote{Other applications of isoperimetric inequality to thermodynamics are found in Refs.~\cite{FrimGeometricBound2022,MiangolarraGeometryOfFiniteTime2022}.}.
\nocite{FrimGeometricBound2022,MiangolarraGeometryOfFiniteTime2022}

Below, we illustrate our result with a theoretical and a practical application. As a theoretical application, we relate $\asymm$ to the eigenvalues of the rate matrix. This leads to a proof of a thermodynamic bound on the coherence of noisy oscillations, which was previously conjectured by Barato and Seifert~\cite{baratoCoherence2017}. As a practical application, we show that chemical driving force bounds directed information flow in biochemical signal transduction.

\paragraph{System and formulation} 
Here we describe our physical setup and define the quantities that appear in our result. We consider a stochastic system modeled as a Markov jump system with a finite number of mesoscopic states $i\in\{1,2,\dots,n\}$, where the transition rate from state $j$ to $i$ is $R_{ij}$ for $i\neq j$. We define the rate matrix $\R \equiv (R_{ij})$ by filling the diagonal elements with $R_{ii} = - \sum_{k : k\neq i} R_{ki}$. The dynamics of the probability distribution $\bm{p}=(p_1,\dots,p_n)^{\mathsf{T}}$ obeys $d\bm{p}/dt=\R\bm{p}$. We use  $\bm\st$ to indicate a steady-state distribution satisfying $\R\bm{\st}=\bm 0$. For convenience, we use  $\flux_{ij}=R_{ij} \st_j$ to indicate the one-way steady-state flux from $j$ to $i$ (with diagonals $\flux_{ii}=R_{ii}\st_i\leq 0$).

We define observables $a$ and $b$ to be any functions of the states, denoted by $a_i$ and $b_i$.
In steady state, their cross-correlation at time lag $\tau$ can be written as
\begin{equation}
    \label{eq:CbaDef}
    C_{ba}^\tau 
    = \sum_{i,j}(e^{\tau\R })_{ij} \st_j b_i a_j 
    \approx  \sum_{i}\st_i  b_i a_i + \tau \sum_{i,j} \flux_{ij} b_i a_j ,
\end{equation}
where $\approx$ means equality to first order in $\tau$, which follows by expanding the exponential.   
The normalization terms in Eq.~\eqref{eq:asymm_def} can be written as 
\begin{equation}
    \Delta_\tau C_{aa}\approx
    -\tau \sum_{i,j} \flux_{ij} a_i a_j=
    \frac{\tau}{2}\sum_{i,j} \flux_{ij}(a_i - a_j)^2 ,
    \label{eq:dCaaDef}
\end{equation}
which follows from $\sum_i \flux_{ij} = \sum_j \flux_{ij} = 0$.
Plugging into \eqref{eq:asymm_def} gives
\begin{equation}
   \asymm =  \frac{\sum_{i , j} \flux_{ij} ( b_i a_j - b_j a_i)  }{2 \sqrt{-\sum_{i,j}  \flux_{ij} a_i a_j}  \sqrt{-\sum_{i,j} \flux_{ij} b_i b_j} }  .
   \label{eq:asymmExplicit}
\end{equation}

A simple cycle $c=(i_{1} \,{\to}\, i_{2} \,{\to}\, \cdots \,{\to}\, i_{n_{c}} \,{\to}\, i_{1})$ is a closed path of $n_{c}\ge 3$ distinct states with $R_{i_{k+1}i_k} > 0$ for $k \in \{ 1, \dots, n_c\}$ (we use the convention $n_c+1 \equiv 1$). 
According to stochastic thermodynamics~\cite{schnakenberg1976network,seifert2012stochastic}, the cycle affinity $\aff_c$ is related to the transition rates by
\begin{equation}
    \aff_c = \ln\left( \frac{R_{i_2 i_1} R_{i_3 i_2}\cdots R_{i_1 i_{n_c}}}{R_{i_1 i_2} R_{i_2 i_3}\cdots R_{i_{n_c} i_1}} \right)\,.
    \label{eq:cycle_affinity}
\end{equation}

With these definitions, the main result~\eqref{eq:main} holds. This bound can be saturated, even in systems arbitrarily far from equilibrium. In particular, for a unicyclic system consisting of a single cycle $c = (1 \,{\to}\, 2 \,{\to}\, \cdots \,{\to}\, n \,{\to}\, 1)$, the first (tighter) bound holds with equality if and only if the one-way fluxes are translation invariant, i.e., $\flux_{12}=\flux_{23} =\cdots = \flux_{n1}$ and $\flux_{21}=\flux_{32} = \cdots = \flux_{1n}$, and there exists a scaling constant $\gamma $ such that the points $(\gamma a_1, b_1), \dots, (\gamma a_n, b_n)$ form a regular $n$-sided polygon on the $a$--$b$ plane in this order.

\paragraph{Tighter bounds}
Under the same setup, we can prove a tighter version of our result,
\begin{equation}
    \vert \asymm\vert
    \leq \max_{c \in \CS} \frac{n_c\tanh(\aff_{c}/2n_{c})}{n'_c \tan(\pi/n'_{c})} 
    \leq \max_{c \in \CS} \frac{\aff_c/2n_c'}{\tan(\pi/n_c')}
    \leq \max_{c \in \CS}  \frac{\aff_{c}}{2\pi}.
    \label{eq:main_tighter}
\end{equation}
Here, $n'_c$ is the number of times the joint value $(a, b)$ changes over course of cycle $c$, which satisfies $n'_c \leq n_c$. $\CS \coloneqq \Casy \cap \Cuni$ is a restricted set of cycles. $\Casy$ is the set of simple cycles $c = (i_1 \,{\to}\, \cdots \,{\to}\, i_{n_c} \,{\to}\, i_1)$ that satisfy $\sum_{k=1}^{n_c} (b_{i_{k+1}} a_{i_k} - b_{i_k} a_{i_{k+1}}) \neq 0$, namely, simple cycles with nonzero net asymmetry. This restriction means that cycles with zero net asymmetry cannot contribute to $C_{ba}^\tau - C_{ab}^\tau$, no matter how large $\aff_c$ is. $\Cuni$ is a restricted set of cycles generated by the so-called uniform cycle decomposition~\cite{PietzonkaAffinity2016}. This restriction not only makes the inequality tighter, it also provides a fast algorithm for solving the maximization over cycles. 
The simpler version~\eqref{eq:main} is recovered from the first bound in Eq.~\eqref{eq:main_tighter} by using $n'_c \tan(\pi/n'_c) \geq n_c \tan(\pi/n_c)$ and $\max_{c\in\CS} \hspace{0.1em}(\cdot ) \leq \max_c\hspace{0.1em} (\cdot )$.
The second bound in~\eqref{eq:main_tighter}, which follows from $\tanh(\aff_c/2n_c)\le \aff_c/2n_c$, is convenient when $n'_c$ is known but the number of underlying states $n_c$ is unknown.

We can also obtain a tighter result by considering a restricted setup: If the two observables are bipartite, i.e., $a$ and $b$ do not change simultaneously in any transition, then
\begin{equation}
    \vert \asymm\vert
    \leq \max_{c \in \CS} \frac{n_c\tanh(\aff_{c}/2n_{c})}{4} 
    \leq \max_{c \in \CS} \frac{\aff_c}{8}
    \leq \max_{c \in \CS}  \frac{\aff_c}{2\pi}\,.
    \label{eq:main_bipartite}
\end{equation}
These bounds may be even tighter than Eq.~\eqref{eq:main_tighter}.

\paragraph{Application 1:~Coherence of noisy oscillation}
We present a theoretical application of our result by proving a thermodynamic bound on the coherence of noisy oscillation. Noisy oscillations are ubiquitous in biological systems~\cite{FerrellModelingTheCellCycle2011,GoldbeterClocks2008}, 
but the oscillation should be coherent in time for reliable biological functionality~\cite{BarkaiCircadianClocks2000,GaspardCorrelationTime2002}. Coherence is supported at the cost of dissipation, thus the relation between thermodynamic cost and the coherence of noisy oscillations is actively studied~\cite{CaoFreeEnergy2015,baratoCoherence2017,NguyenPhaseTransition2018,OberreiterSubharmonic2019,delJunco2020High,delJunco2020Robust,RemleinCoherenceOfOscillations2022,UhlAffinity2019,OberreiterUniversalMinimalCost2022,kolchinsky2023thermodynamicSpectral}.

The coherence of oscillation is quantified by the number of oscillations that occur before the steady-state autocorrelations die down. 
To introduce this, let $\lambda_{1},\dots,\lambda_{n}$ be the eigenvalues of the rate matrix $\R$ with real and imaginary parts $\lambda_{\alpha}=-\lr_{\alpha}+\imag\li_{\alpha}$. Suppose for simplicity that the rate matrix is diagonalizable, in which case the matrix exponential can be expressed as $e^{\tau\R } = \sum_\alpha \exp(-\lr_\alpha \tau) \exp(\imag \li_\alpha \tau ) \revecA  {\levecA}{}^{\mathsf T}$, where $\levecA$ and $\revecA$ indicate the left and right eigenvectors of $\R$ normalized so that $ \levecA {}^{\mathsf T} \revecA = 1$. Plugging this expression into Eq.~\eqref{eq:CbaDef} shows that any two-time correlation can be written as a sum over the eigenmodes of $\R$, where the contribution from mode $\alpha$ decays with timescale $(\lr_\alpha)^{-1}$ and oscillates with period $2\pi\vert \li_\alpha\vert^{-1}$. 
The number of coherent oscillations for mode $\alpha$ is the ratio between the decay time and the period of the oscillations,  $(\lr_\alpha)^{-1}/2\pi\vert\li_\alpha\vert^{-1}=\vert\li_\alpha\vert/2\pi\lr_\alpha$~\cite{QianPumped2000,baratoCoherence2017}.

Barato and Seifert~\cite{baratoCoherence2017} conjectured, based on numerical evidence, that the slowest decay mode (with the smallest nonzero $\lr_\alpha$) obeys a thermodynamic bound,
\begin{equation}
    \frac{\left\vert \li_\alpha \right\vert}{2\pi\lr_\alpha}
    \leq\max_{c}\frac{\tanh(\aff_{c}/2n_{c})}{2\pi \tan(\pi/n_{c})} .
    \label{eq:Barato_Seifert}
\end{equation}
This bound implies that coherent oscillations require strong thermodynamic driving. Despite its fundamental and profound nature, this inequality has not been rigorously proven.

Here we use our result~\eqref{eq:main} to prove that~\eqref{eq:Barato_Seifert} holds for all modes $\alpha$. We normalize the right eigenvector $\revecA$ so that $\sum_i \vert u^{(\alpha)}_i \vert^{2}/\st_i =1$
and define two observables 
$a_i = \operatorname{Im} u^{(\alpha)}_i/\st_i$ and  $b_i = \Re u^{(\alpha)}_i/\st_i$. 
Using Eqs.~\eqref{eq:CbaDef} and \eqref{eq:dCaaDef}, the imaginary and real parts of $\lambda_\alpha$ can be written as
\begin{equation}
    \li_\alpha = \lim_{\tau \to 0} \frac{ C_{ba}^\tau- C_{ab}^\tau}{\tau}, \quad 
    \lr_\alpha = \lim_{\tau \to 0 } \frac{\Delta_\tau C_{aa} + \Delta_\tau C_{bb}}{\tau} ,
\label{eq:eigenvalue}
\end{equation}
as derived in the Supplemental Material~\cite{SM}.
\nocite{radoIsoperimetricInequalityLebesgue1947,radoLemmaTopologicalIndex1936,scottFamilyPolygons1982,faryIsoperimetryVariableMetric1982,fanAlgebraicProofIsoperimetric1955}
Combining Eqs.~\eqref{eq:eigenvalue} and \eqref{eq:main}, together with the  inequality $(x+y)/2 \geq \sqrt{xy}$, proves the conjecture~\eqref{eq:Barato_Seifert}. The bound~\eqref{eq:Barato_Seifert} is saturated in unicyclic systems with uniform transition rates for the slowest decay mode.

To our knowledge, the expression~\eqref{eq:eigenvalue} is new to the literature. More generally, our analytical approach to the eigenvalues of rate matrices complements classical results on this topic~\cite{GershgorinAbgrenzung1931,dmitrievCharacteristicNumbers1945,dmitrievCharacteristicRoots1946,KarpelevichCharacteristicRoots1951,SwiftMasterThesis1972,HornMatrixAnalysis2012}, and it may contribute to ongoing research on the relationship between thermodynamics and complex eigenvalues~\cite{OberreiterSubharmonic2019,delJunco2020High,delJunco2020Robust,RemleinCoherenceOfOscillations2022,UhlAffinity2019,OberreiterUniversalMinimalCost2022,kolchinsky2023thermodynamicSpectral}. For instance, Oberreiter \textit{et al.}~\cite{OberreiterUniversalMinimalCost2022} recently conjectured another thermodynamic bound on $\vert\li_\alpha\vert/2\pi\lr_\alpha$ in terms of entropy production rate. Although our approach alone cannot directly prove this newer conjecture, it may be useful when combined with other ideas.

The result~\eqref{eq:Barato_Seifert} also has practical implications, as it may be used to infer cyclic affinity from empirical observations, assuming some two-time correlation function exhibits a clear damped oscillation corresponding to a particular mode. In this case, one would not directly measure the ``observables'' $a$ and $b$, but rather estimate $\li_\alpha$ and $\lr_\alpha$  by fitting the two-time correlation function with a damped oscillation.

\begin{figure}[t]
    \centering
    \includegraphics[width=\hsize,clip]{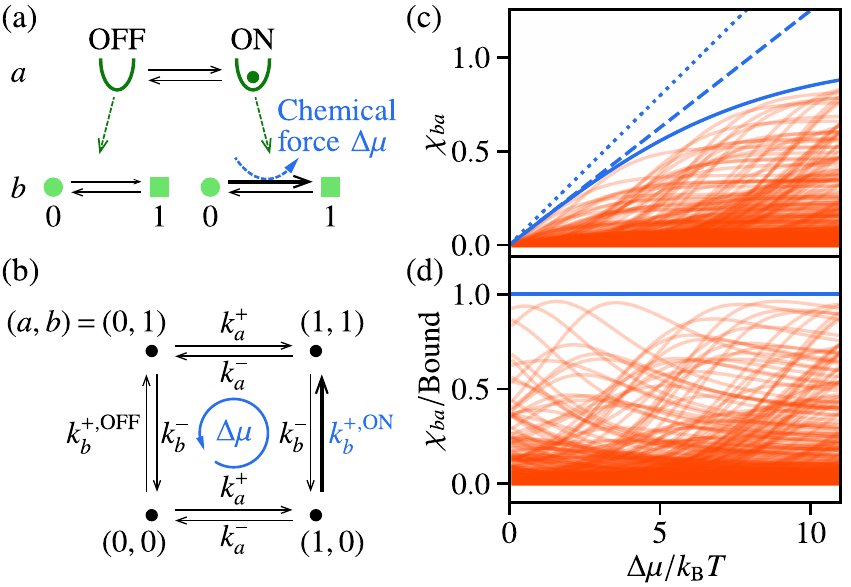}
    \caption{(a) Simple model of biological signal transduction~\cite{MehtaEnergeticCosts2012}. 
    (b) Formulation as a Markov jump system with a nonequilibrium cycle. 
    (c) Validation of our bounds. 
    Orange indicates $\asymm$ for varying chemical force $\Delta \mu$, which determines the rate of the transition $k_b^{+,\mathrm{ON}}$ (other kinetic rates set to random but fixed values). $\chi_{ba}$ is nonnegative for $\Delta\mu \geq 0$ in this model. Blue indicates the three upper bounds from Eq.~\eqref{eq:receptor}. 
    (d) The ratio between $\chi_{ba}$ and the tightest bound $\tanh(\Delta\mu/8\kBphys T)$ in Eq.~\eqref{eq:receptor}. }
    \label{fig_application}
\end{figure}

\paragraph{Application 2:~Signal transduction in a biochemical system} 
One of the goals of stochastic thermodynamics is to understand the costs of information processing in biochemical systems~\cite{MehtaEnergeticCosts2012,lan2012energy,barato2013information,ItoInformationThermodynamics2013,govern2014energy,govern2014optimal,barato2014efficiency,sartori2014thermodynamic,ten2016fundamental,ito2015maxwell,mehta2016landauer}. To illustrate a practical application of our result, we derive a thermodynamic bound on directed information flow in a standard model of biochemical signal transduction [Fig.~\ref{fig_application}(a)]~\cite{MehtaEnergeticCosts2012}.

The model consists of an upstream receptor and a downstream protein. The upstream receptor stochastically switches between ``OFF'' and ``ON'' states due to ligand binding, corresponding to the observable $a=0,1$. The downstream stochastically switches between ``0'' (inactive) and ``1'' (active) states, corresponding the observable $b=0,1$. When the upstream is ON, the activation of the downstream ($0 \,{\to}\, 1$) is driven by a chemical force $\Delta\mu > 0$. For example, if the driving is provided by the hydrolysis of a molecule of adenosine triphosphate (ATP), $\Delta\mu = \mu_\mathrm{ATP} - \mu_\mathrm{ADP} - \mu_\mathrm{Pi}$, with $\mu_\mathrm{X}$ being the environmental chemical potential of $\mathrm{X}$. The bipartite dynamics are modeled as a four-state Markov jump system depicted in Fig.~\ref{fig_application}(b). The unique cycle in this system has cycle affinity $\aff = \Delta \mu /\kBphys T$, where $\kBphys$ is the Boltzmann constant and $T$ is the environmental temperature. 

In this model, the cross-correlation $C_{ba}^\tau$ is the joint probability of the receptor being ON at time $t$ and the protein being active at time $t+\tau$, and vice versa for $C_{ab}^\tau$. Therefore, the asymmetry $C_{ba}^\tau- C_{ab}^\tau$ provides a simple and natural measure of directed information flow from $a$ to $b$~\cite{NolteIdentifyingInteractions2006,SisanEventOrdering2010,BorysovCrossCorrelationAsymmetries2014}. 
As for the normalization factor, Eq.~\eqref{eq:dCaaDef} implies that $\Delta_\tau C_{aa}$ is one half of the expected number of switching events of $a$ during a short period $\tau$, and similarly for $\Delta_\tau C_{bb}$. Therefore, $\asymm$ is normalized by the frequency of the switches of $a$ and $b$.

Our general result~\eqref{eq:main_tighter} specializes to
\begin{equation}
    \vert \asymm\vert  \leq \tanh \frac{\Delta\mu}{8 \kBphys T} \le \frac{\Delta\mu}{ 8 \kBphys T} \le \frac{\Delta\mu}{ 2\pi \kBphys T}.
    \label{eq:receptor}
\end{equation}
See Figs.~\ref{fig_application}(c) and \ref{fig_application}(d). Thus, the chemical force $\Delta\mu$ bounds directed information flow, irrespective of the details of the transition rates. Even if the transition rates are perturbed, the bound is not affected as long as  $\Delta\mu$ is unchanged. 

Although Eq.~\eqref{eq:receptor} was motivated by a specific model of signal transduction~\cite{MehtaEnergeticCosts2012}, the result is much more general. The bound $\vert \asymm\vert  \leq {\Delta\mu}/{ 2\pi \kBphys T}$ applies to any signal transduction system that includes a binary upstream and downstream, including multicyclic systems and systems with nonobserved transition and states, as long as the maximum cycle affinity is $\Delta \mu / \kBphys T$. The tighter bound $\vert \asymm\vert \leq {\Delta\mu}/{ 8 \kBphys T}$ holds when, in addition, the upstream and downstream observables are bipartite (do not change at the same time), as follows from Eq.~\eqref{eq:main_bipartite}.

\paragraph{Derivation}
We sketch the derivation of Eq.~\eqref{eq:main}, restricting our attention to unicyclic systems for simplicity. Further details, including derivation of our tighter bound~\eqref{eq:main_tighter} and consideration of multicyclic systems, are in the Supplemental Material~\cite{SM}.

For each transition $j \,{\to}\, i$, we define the net probability current $\cur_{ij}= \flux_{ij} -  \flux_{ji}$ and the dynamical activity $\act_{ij} =  \flux_{ij} + \flux_{ji}$. We also define $\Omega_{ij} \coloneqq \frac 12 (b_i a_j  - b_j a_i ) $ and $L_{ij} \coloneqq \sqrt{ (a_i-a_j)^2 + (b_i - b_j)^2 }$. 

We recast $\chi_{ba}$ in a convenient form. For a unicyclic system with $c=(1\,{\to}\, 2 \,{\to}\, \cdots \,{\to}\, n \,{\to}\, 1)$, the steady-state currents are uniform, $\cur_{21} = \cur_{32} = \cdots = \cur_{1n} \eqqcolon \cur $. We assume without loss of generality that $\cur \ge 0$ (otherwise, we may consider the cycle in reverse). 
Next, since $\vert \asymm \vert$ is invariant under multiplication of $a$ and $b$ by any pair of real numbers, we may assume without loss of generality that $a$ and $b$ are scaled to satisfy $ \sum_{i,j} \flux_{ij} a_ia_j = \sum_{i,j} \flux_{ij} b_ib_j $. Combining this assumption with Eq.~\eqref{eq:dCaaDef}, we may rewrite the denominator of Eq.~\eqref{eq:asymmExplicit} as $- \sum_{i,j} \flux_{ij} ( a_i a_j + b_i b_j ) = \frac 12 \sum_{i>j} \act_{ij} L_{ij}^2 $. This gives  
\begin{equation}
    \asymm  = \frac{4\cur \sum_i \Omega_{i+1,i} }{ \sum_i \act_{i+1,i} L_{i+1,i}^2 }\,.
    \label{eq:gg}
\end{equation}

We use two techniques to bound the right side of Eq.~\eqref{eq:gg}. First, we generalize the short-time TUR~\cite{otsubo2020estimating} to
\begin{equation}
    \frac{(\cur\sum_i L_{i+1,i})^{2}}{\sum_i \act_{i+1,i}L_{i+1,i}^{2}}
    \leq \cur n \tanh \frac{\aff}{2n} \,,
\label{eq:genTUR}
\end{equation}
where $\mathcal{F}$ is the cycle affinity of the unique cycle.
To prove Eq.~\eqref{eq:genTUR}, we first use the Cauchy--Schwarz inequality to show that the left-hand side is less than $\cur^2 \sum_i \act_{i+1,i}^{\,-1}$. Next, we rewrite the affinity as $\aff = 2n \sum_i n^{\,-1} \artanh (\cur/\act_{i+1,i})$ and apply Jensen's inequality to $\artanh$ to show that $ \cur \sum_i (\cur/\act_{i+1,i})$ is less than the right-hand side.

The second technique uses a geometrical interpretation of $\Omega_{ij}$ and $L_{ij}$. 
$\Omega_{i+1,i}$ is the signed area swept by the observables during the transition $(a_i,b_i) \,{\to}\, (a_{i+1},b_{i+1})$, while 
$L_{i+1,i}$ is the length of this transition  [Fig.~\ref{fig_illustration}(d)]. Over the course of the cycle, the total signed area is $\sum_i  \Omega_{i+1,i}$ and the length of the curve is 
$\sum_i L_{i+1,i} $. We relate the area and length using the isoperimetric inequality,
\begin{equation}
    \qty( 4n\tan\frac{\pi}{n} ) \, \Bvert {\sum_i  \Omega_{i+1,i} } \leq \Bparen{ \sum_i L_{i+1,i} } ^{2}. \label{eq:isoperimetric}
\end{equation}
As shown in the Supplemental Material~\cite{SM}, this inequality holds even if the curve has self-intersections. Combining Eqs.~\eqref{eq:gg}--\eqref{eq:isoperimetric} leads to our main result.

\paragraph{Discussion}
In this Letter, we uncovered a universal thermodynamic bound on the asymmetry of cross-correlation between observables. This result holds for any pair of observables in a finite-state stochastic system in steady state. It is experimentally accessible, relying only on short-time two-point correlation functions. 

Our result is similar in spirit to TURs, which also relate the statistical and thermodynamic properties of nonequilibrium steady states~\cite{Barato2015thermodynamic,GingrichInferringDissipation2017,SeifertFromStochasticThermodynamics2019,horowitz2020thermodynamic}. However, the two approaches differ both in statistical and thermodynamic aspects. First, our bound is defined using 
two-time correlations of 
state observables (e.g., counts of chemical species, voltages, etc.), while TURs are usually defined using 
the mean and variance of 
antisymmetric current observables (e.g., chemical reaction fluxes, electric currents, etc.).  
Although the asymmetry $C_{ba}^\tau-C_{ab}^\tau$ can be interpreted as the mean of a specific antisymmetric current observable, the variance of this observable lacks an intuitive physical or statistical interpretation.    
Second, our bound uses the cycle affinity as the measure of thermodynamic cost, while TURs use the entropy production rate. The affinity is determined by macroscopic parameters (such as environmental chemical potentials) and does not depend on the steady-state distribution. Therefore, it can be interpreted and manipulated at the macroscopic level, and it is robust against microscopic perturbations. In contrast, 
the entropy production rate captures the resulting dissipation rate and is sensitive to microscopic details. Thus, these two measures provide different and complementary characterizations of the thermodynamic cost.
To our knowledge, there is no way to use TURs to bound cycle affinities in general (multicyclic) systems~\footnote{Inequalities on cycle affinities in terms of cycle-based statistics~\cite{PolettiniTightUncertaintyRelation2022} and waiting-time statistics~\cite{VanDerMeerThermodynamicInference2022} have recently been proposed. However, these inequalities cannot be used to study the asymmetry of cross-correlations.}. 
\nocite{PolettiniTightUncertaintyRelation2022,VanDerMeerThermodynamicInference2022}

It is interesting to consider our bound~\eqref{eq:main} for finite time lags without taking the short-time limit $\lim_{\tau\to0}$ in Eq.~\eqref{eq:asymm_def}. Given  
numerical evidence presented in the Supplemental Material~\cite{SM}, we conjecture that our bound holds for all $\tau$. Proving this conjecture is an interesting direction for future work.
In addition, future work may generalize our approach to cases when cross-correlations between three or more observables are available at once. Finally, it would be interesting to extend our analysis to continuous-state systems and quantum systems.

\vspace{\sectionsep}
\begin{acknowledgements}
We thank Kohei Yoshimura for discussions. N.~O.~is supported by JSPS KAKENHI Grant No.~23KJ0732. S.~I.~is supported by JSPS KAKENHI Grants No.~19H05796, No.~21H01560, No.~22H01141, and No. 23H00467, JST ERATO-FS Grant No.~JPMJER2204, and UTEC-UTokyo FSI Research Grant Program.
\end{acknowledgements}

\onecolumngrid

\appendix
\counterwithout{equation}{section}

\clearpage

\onecolumngrid

\setcounter{equation}{0}
\setcounter{figure}{0}
\setcounter{table}{0}
\setcounter{page}{1}

\renewcommand{\theequation}{S\arabic{equation}}
\renewcommand{\thefigure}{S\arabic{figure}}
\renewcommand{\thesection}{\Alph{section}}
\renewcommand{\thesubsection}{\arabic{subsection}}

\let\oldsection\section
\let\oldsubsection\subsection
\let\olduppercase\uppercase
\renewcommand{\section}[1]{%
    \refstepcounter{section}
    \oldsection{\thesection.~~~#1}
}
\renewcommand{\subsection}[1]{%
    \refstepcounter{subsection}
    \oldsubsection{\thesection.\thesubsection~~~#1}
}

\titleformat*{\section}{\centering\fontsize{10.5pt}{\baselineskip}\selectfont\bfseries}
\titleformat*{\subsection}{\centering\fontsize{10.5pt}{\baselineskip}\selectfont\bfseries}
\titleformat*{\subsubsection}{\centering\fontsize{10.5pt}{\baselineskip}\selectfont\itshape}

\titlespacing*{\section}{\linewidth}{3em}{0.4em}
\titlespacing*{\subsection}{\linewidth}{1.5em}{0.4em}
\titlespacing*{\subsubsection}{\linewidth}{1.5em}{0.4em}

\fontsize{10.5pt}{12.0pt}\selectfont
\fontdimen2\font=2.625pt

\begin{center}
\textbf{\large Supplemental Material for\\[0.1em]
\MyTitle}
\vspace*{1em}

Naruo\hspace{0.35em}Ohga,\hspace{0.35em}Sosuke\hspace{0.35em}Ito,\hspace{0.35em}and\hspace{0.35em}Artemy\hspace{0.35em}Kolchinsky
\vspace*{1.5em}
\end{center}

In Sec.~\ref{sec:eigenvalue}, we prove the expression of eigenvalues in Eq.~\eqref{eq:eigenvalue}. 
In Sec.~\ref{sec:derivation}, we provide the full derivation of our main results and the tighter versions, Eqs.~\eqref{eq:main}, \eqref{eq:main_tighter} and \eqref{eq:main_bipartite}, for general multicyclic systems.
In Sec.~\ref{sec:finitetime}, we show numerical evidence that supports the finite-$\tau$ version of our bound.
In Sec.~\ref{sec:additional}, we provide supplemental discussions, such as implicit assumptions made in the main text and the equality conditions. 

\vspace*{1.5em}

\twocolumngrid

\section{Proof of the expressions of eigenvalues in Eq.~\texorpdfstring{\eqref{eq:eigenvalue}}{(10)}}
\label{sec:eigenvalue}

We prove the expression of $\lr_\alpha$ and $\li_\alpha$ by two-time correlations in Eq.~\eqref{eq:eigenvalue}, which applies to any of the eigenvalues. For conciseness, we fix an $\alpha$ and omit the suffix $\alpha$ in $\lambda_\alpha$ and $u_i^{(\alpha)}$ below. 

\paragraph{Proof Eq.~\eqref{eq:eigenvalue}}
Thanks to the special normalization of the eigenvector, $\sum_i \vert u_i\vert^2 /q_i = 1$, we can rewrite the eigenvalue $\lambda$ as
\begin{equation}
    \lambda 
    = \sum_i \frac{u_i^*}{q_i} \lambda u_i 
    = \sum_{i,j} \frac{u_i^*}{q_i} R_{ij} u_j
    = \sum_{i,j} R_{ij} q_j \frac{u_i^*}{q_i} \frac{u_j}{q_j} .
    \label{sm:lambda}
\end{equation}
Using the definition $\flux_{ij}=R_{ij} q_j$ and the observables $a_i = \Im u_i / q_i$ and $b_i = \Re u_i / q_i$ introduced in the main text, we obtain
\begin{align}
    \lambda 
    &= \sum_{i,j} \flux_{ij} (b_i - \imag a_i) (b_j + \imag a_j),
    \notag \\
    &=  \sum_{i,j} \flux_{ij} (b_i b_j + a_i a_j + \imag b_i a_j - \imag a_i b_j ).
\end{align}
Therefore, the real and imaginary parts, $\lambda = -\lr + \imag \li$, are given by
\begin{align}
    \li 
    &= \sum_{i,j} \flux_{ij} (b_i a_j - a_i b_j ) 
    = \lim_{\tau \to 0} \frac{C_{ba}^\tau - C_{ab}^\tau}{\tau} ,
    \label{sm:li_final} \\
    \lr
    &= - \sum_{i,j} \flux_{ij} (b_i b_j + a_i a_j)
    = \lim_{\tau \to 0} \frac{\Delta_\tau C_{aa} + \Delta_\tau C_{bb}}{\tau},
    \label{sm:lr_final}
\end{align}
where we have used the expressions of the correlations in Eqs.~\eqref{eq:CbaDef} and \eqref{eq:dCaaDef}. \qed

We remark that we can also recast Eq.~\eqref{sm:lambda} in terms of the currents $\cur_{ij} = \flux_{ij} - \flux_{ji}$ and the dynamical activity $\act_{ij} = \flux_{ij} + \flux_{ji}$ as
\begin{equation}
     \li 
     =  \sum_{i>j} \cur_{ij} \Im \qty( \frac{u^*_iu_j}{q_iq_j} ) , \quad 
     \lr 
     = \frac 12 \sum_{i>j} \act_{ij} \Bvert{ \frac{u_i}{q_i} - \frac{u_j}{q_j} }^2 ,
\end{equation}
where we have used $\sum_i \flux_{ij} = \sum_j \flux_{ij} = 0$ to obtain the expression of $\lr$. This expression connects the system's global dynamics characterized by the eigenvalues to the local dynamics characterized by $\cur_{ij}$ and $\act_{ij}$: The global oscillation $\li$ due to a current along a cycle is governed by the local current $\cur_{ij}$, and the global relaxation $\lr$ is governed by the local frequency of transitions $\act_{ij}$.

\section{Derivation of the main results}
\label{sec:derivation}

We show the complete proof of our main result~\eqref{eq:main} and the tighter versions, Eqs.~\eqref{eq:main_tighter} and \eqref{eq:main_bipartite}. 
We use the symbols $\cur_{ij}$, $\act_{ij}$, $\Omega_{ij}$, and $L_{ij}$ defined in the \textit{Derivation} section in the main text.

The derivation proceeds in the same way as in the derivation for unicyclic systems. We rewrite the measure of asymmetry $\chi_{ba}$ in Secs.~\ref{subsec:rewrite}--\ref{subsec:uniform_decomp}, develop two tools in Secs.~\ref{subsec:genTUR} and \ref{subsec:isoperimetric}, and finally prove Eqs.~\eqref{eq:main}, \eqref{eq:main_tighter}, and \eqref{eq:main_bipartite} in Sec.~\ref{subsec:final}.

\subsection{Rewriting the ratio \texorpdfstring{$\chi_{ba}$}{chi\_ba}}
\label{subsec:rewrite}

The first step of the proof is to rewrite $\chi_{ba}$ in a convenient form, as has been done in the main text. Since $\vert \asymm \vert$ is invariant under multiplication of $a$ and $b$ by any pair of real numbers, we may assume without loss of generality that $a$ and $b$ are scaled so that
\begin{equation}
    \sum_{i,j} \flux_{ij} a_ia_j = \sum_{i,j} \flux_{ij} b_ib_j . 
    \label{sm:rescale_assumption}
\end{equation}
Under this assumption, we can rewrite $\chi_{ba}$ in Eq.~\eqref{eq:asymmExplicit} as
\begin{align}
    \chi_{ba} &= \frac{\sum_{i,j} \flux_{ij} (b_i a_j - b_j a_i)}{- \sum_{i,j } \flux_{ij} ( a_i a_j  + b_i b_j) }
    \notag \\
    &= \frac{ \sum_{i,j} \flux_{ij} (b_i a_j - b_j a_i) }{ \frac 12 \sum_{i,j} \flux_{ij} \qty[ (a_i - a_j)^2 + (b_i - b_j )^2 ] }
    \notag \\
    &= \frac{4 \sum_{i,j} \flux_{ij} \Omega_{ij} }{ \sum_{i,j} \flux_{ij} L^2_{ij}  }
    \notag \\
    &= \frac{4 \sum_{i>j} \cur_{ij} \Omega_{ij} }{ \sum_{i>j} \act_{ij} L^2_{ij}  } \,,
\end{align}
where the second equality is due to the second equality in Eq.~\eqref{eq:dCaaDef}, and the last equality follows from $\Omega_{ij}=-\Omega_{ji}$ and $L_{ij} = L_{ji}$. The absolute value reads
\begin{equation}
    \vert \chi_{ba} \vert = \frac{4 \bigl\vert \sum_{i>j} \cur_{ij} \Omega_{ij} \bigr\vert }{ \sum_{i>j} \act_{ij} L^2_{ij}  }\, ,
    \label{sm:chi_convenient}
\end{equation}
which follows from $\act_{ij} \geq 0$ and $L_{ij} \geq 0$.

\subsection{Notation}

We introduce notation suitable for analyzing multicyclic systems. We use the symbol $e=(j\,{\to}\, i)$ to denote the transition from $j$ to $i$, which we call a (directed) edge after the terminology in graph theory. For an edge $e=(j\,{\to}\, i)$, we use the symbol $-e$ to denote the transition in the opposite direction $-e=(i \,{\to}\, j)$, and we use $\flux_e \equiv \flux_{ij}$, $\flux_{-e} \equiv \flux_{ji}$, $\cur_e \equiv \cur_{ij}$, $\act_e \equiv \act_{ij}$,
$\Omega_e \equiv \Omega_{ij}$, and $L_e \equiv L_{ij}$.
We define the set of edges with a positive current, $\Eplus \coloneqq \Set{e|\cur_e>0}$.

We regard a cycle $c=(i_{1} \,{\to}\, i_{2} \,{\to}\, \cdots \,{\to}\, i_{n_{c}} \,{\to}\, i_{1})$ as a sequence of edges $(e_1, e_2,\dots, e_{n_c})$ with $e_k = (i_{k} \,{\to}\, i_{k+1})$. The summation over the cycle reads $\sum_{e \in c} X_e \coloneqq \sum_{k=1}^{n_c} X_{e_k}$ for any quantity $X_e$ associated with the edges. In particular, the cycle affinity $\aff_c$ in Eq.~\eqref{eq:cycle_affinity} is rewritten as
\begin{equation}
  \aff_c
  = \sum_{k=1}^{n_c} \ln\frac{\flux_{i_{k+1}i_k}}{\flux_{i_k i_{k+1}}} 
  = \sum_{e\in c} \ln \frac{\flux_{e}}{\flux_{-e}}\,.
  \label{sm:cycle_affinity}
\end{equation}

\subsection{Uniform cycle decomposition}
\label{subsec:uniform_decomp}

In the steady state, the currents $\cur_e$ should not produce any net change of the probability at each state, and thus the currents are written as a superposition of currents that circulate around the cycles. This idea is called the \textit{cycle decomposition} or \textit{Schnakenberg decomposition}~\cite{schnakenberg1976network}. 

In particular, Ref.~\cite{PietzonkaAffinity2016} shows that there exists a special decomposition with good properties called a \textit{uniform cycle decomposition}. A uniform cycle decomposition employs a subset of simple cycles $\Cuni$ whose every cycle is aligned with the edge currents, i.e., any edge $e\in c $ for every cycle $c\in\Cuni$ belongs to $\Eplus$. Then, it decomposes the edge currents $\cur_e > 0$ for $e\in\Eplus$ into positive cycle currents $\cur_c > 0$ for $c\in \Cuni$ as
\begin{equation}
    \cur_{e}=\sum_{c \in \Cuni} S_{ec}\cur_{c},
    \label{sm:uniform_decomp}
\end{equation}
where $S_{ec}=1$ for $e\in c$ and $S_{ec}=0$ for $e\notin c$. Note that $-e\notin c$ for any $e\in \Eplus$ and $c \in \Cuni$.

Reference \cite{PietzonkaAffinity2016} discusses an algorithm to find $\Cuni$ and $\cur_c$. It assigns the edge currents to cycle currents by iterating the following steps: (1) find an edge $e^*$ with the minimum unassigned edge current; (2) pick a cycle $c$ that passes through the edge $e^*$ and is aligned with the edge currents, whose existence is guaranteed; add $c$ to $\Cuni$; (3) set the cycle current $\cur_{c}= \cur_{e^*}$; (4) for all $e\in c$, subtract $\cur_c$ from the edge currents $\cur_e$ and set $S_{ec}=1$, while for all other edges $e' \notin c$, set $S_{e'c}=0$. Steps (1)--(4) are repeated until all edge currents are assigned to cycle currents. Although the construction of $\Cuni$ and $\cur_c$ is not necessarily unique, below we fix one out of them.

For any quantity $X_e$ associated with the edges $e\in \Eplus$, we obtain 
\begin{align}
    \sum_{e \in \Eplus} \cur_e X_e 
    &= \sum_{c\in\Cuni}  \cur_c \, \Bparen{ \sum_{e \in \Eplus}  S_{ec} X_e }
    \notag \\
    &= \sum_{c \in \Cuni} \cur_c \, \Bparen{\sum_{e\in c} X_e }\,.
    \label{sm:decomp_X}
\end{align}
We also have $\aff_c > 0$ for $c\in\Cuni$ because, in Eq.~\eqref{sm:cycle_affinity}, the sign of $\ln(\flux_e / \flux_{-e})$ is the same as the sign of $\flux_e - \flux_{-e} = \cur_e > 0$.

Using a uniform cycle decomposition and Eq.~\eqref{sm:decomp_X}, we can rewrite the numerator of $\vert \chi_{ba}\vert$ from  Eq.~\eqref{sm:chi_convenient} as
\begin{align}
    \Bvert{ \sum_{i>j} \cur_{ij} \Omega_{ij} }
    &= \Bvert{ \sum_{e \in \Eplus} \cur_e \Omega_e }
    \notag \\
    &= \Bvert{ \sum_{c\in \Cuni} \cur_c \,\Bparen{ \sum_{e\in c} \Omega_e } }
    \notag \\
    &= \Bvert{ \sum_{c\in \CS} \cur_c \,\Bparen{ \sum_{e\in c} \Omega_e } }
    \notag \\
    &\leq \sum_{c\in \CS} \cur_c \, \Bvert{\sum_{e\in c} \Omega_e } \,.
    \label{sm:numerator} 
\end{align}
In the third equality, we narrowed the range of the sum from $\Cuni$ to $ \CS = \Casy \cap \Cuni =  \Set{ c \in \Cuni | \sum_{e\in c} \Omega_e \neq 0 } $, which is the subset of $\Cuni$ with the cycles with nonzero net asymmetry, as introduced in the main text. In the fourth line, we used the triangle inequality and $\cur_c>0$. Similarly, the denominator of $\vert\chi_{ba}\vert$ from Eq.~\eqref{sm:chi_convenient} is rewritten as
\begin{align}
    \sum_{i>j}\act_{ij}L_{ij}^2
    & \geq\sum_{e\in \Eplus}\act_{e} L^2_{e}
    \notag \\ 
    & = \sum_{e\in \Eplus} \cur_{e}\frac{\act_{e}}{\cur_{e}} L^2_{e} 
    \notag \\
    & = \sum_{c\in \Cuni} \cur_{c} \,\Bparen{\sum_{e\in c} \frac{\act_{e}}{\cur_{e}} L^2_{e} }  
    \notag \\
    & \geq \sum_{c\in \CS} \cur_{c} \,\Bparen{ \sum_{e\in c} \frac{\act_{e}}{\cur_{e}} L^2_{e} }\,.
    \label{sm:denomenator}
\end{align}
We have dropped the edges with $ \cur_{ij} = \cur_{ji} = 0 $ in the first inequality and further narrowed the range of the sum in the last inequality using that $\cur_e > 0$ and $\cur_c > 0$ for $e\in c\in \Cuni$. In the following, we develop two tools to evaluate these cycle-wise quantities.

\subsection{Generalized TUR}
\label{subsec:genTUR}

The first tool we will use to derive our result is a generalization of the short-time TUR.
Compared to the original short-time TUR~\cite{otsubo2020estimating}, our generalization uses affinity rather than the entropy production rate, and it deals with each cycle separately.

\paragraph{Lemma}
For any cycle $c \in \Cuni$ from a uniform cycle decomposition, and for any quantity $X_{e}$ associated with the edges $e\in c$,
\begin{equation}
    \frac{\left(\sum_{e\in c} X_e\right)^{2} }{\sum_{e\in c} X_e^2 {\act_{e}}/{\cur_{e}}}
    \leq
    n_c \tanh \qty( \frac{\aff_c}{2n_c} )\,.
    \label{sm:genTUR_multicyclic}
\end{equation}
This inequality is saturated if and only if $\cur_e / \act_e$ and $X_e$ are each constant across the cycle. 

\paragraph{Proof}
We use the Cauchy--Schwarz inequality between
vectors $\qty(X_{e} \!\sqrt{ {\act_{e}} / {\cur_{e}} } ) {}_{e \in c}$ and $\qty( \!\sqrt{{\cur_{e}}/{\act_{e}} } ) {}_{e \in c}  $ to obtain
\begin{equation}
    \frac{\left(\sum_{e\in c} X_e\right)^{2} }{\sum_{e\in c} X_e^2 {\act_{e}}/{\cur_{e}}}
    \leq
    \sum_{e\in c} \frac{\cur_{e}}{\act_{e}},
    \label{sm:CauchySchwarz}
\end{equation}
where we have used that $\cur_e > 0$ for $e \in c$ due to the definition of uniform cycle decomposition.
To relate the right-hand side with the cycle affinity, we rewrite the cycle affinity in Eq.~\eqref{sm:cycle_affinity} as
\begin{align}
    \aff_{c} 
    & = \sum_{e\in c} \ln\frac{\flux_e}{\flux_{-e}} 
    \notag \\
    & = \sum_{e\in c} \ln\frac{\act_{e}+\cur_{e}}{\act_{e}-\cur_{e}}
    \notag \\
    & =2n_{c}\sum_{e\in c} \frac{1}{n_{c}} \artanh \qty(\frac{\cur_{e}}{\act_{e}} ).
\end{align}
Since $\cur_{e} / \act_e \geq 0$, we can apply Jensen's inequality to the $\artanh$ function in the last line to give
\begin{equation}
  \aff_c 
  \geq 2n_{c} \artanh \,\Biggl(\sum_{e\in c} \frac{1}{n_{c}} \frac{\cur_{e}}{\act_{e}} \Biggr).
\end{equation}
Inverting this inequality gives 
\begin{equation}
  \frac{1}{n_{c}} \sum_{e\in c} \frac{\cur_{e}}{\act_{e}}
  \leq  \tanh \qty( \frac{\aff_c}{2n_{c}} ).
\end{equation}
Combining with Eq.~\eqref{sm:CauchySchwarz} gives the desired result. 

The equality condition for the Cauchy--Schwarz inequality is that the vector $\qty(X_e \! \sqrt{\act_e / \cur_e}) {}_{e\in c}$ is proportional to $\qty( \! \sqrt{\cur_e / \act_e} ) {}_{e\in c}$, or equivalently, $X_e \act_e / \cur_e$ is uniform along the cycle. The equality condition for the Jensen's inequality is that $\cur_e / \act_e$ is uniform across the cycle. Rearranging these two conditions gives the equality condition in the statement. \qed

Note that, when the cycle contains an edge with absolute irreversibility, i.e., $R_{ij} > 0 $ and $R_{ji}=0$, the affinity is $\aff_c = \infty$. Equation~\eqref{sm:genTUR_multicyclic} still holds formally for such cases.

\subsection{Isoperimetric inequality}
\label{subsec:isoperimetric}

Next, we prove the other element of the proof, the isoperimetric inequality for polygons. This is a purely mathematical theorem that does not rely on the context. Although the isoperimetric inequality is a well-known result~\cite{OssermanIsoperimetricInequality1978}, we need to generalize the inequality for possibly self-intersecting polygons, which has not been explicitly stated in literature.

\paragraph{Theorem}
For any sequence of $n$ points $(a_1, b_1), \allowbreak \dots,\allowbreak (a_n, b_n)$ on $\mathbb R^2$,  
\begin{equation}
    \qty(4n \tan \frac{\pi}{n} )\,
    \Bvert{ \sum_{i=1}^n \Omega_{i+1,i} }
    \leq \Bparen{ \sum_{i=1}^n  L_{i+1,i} } ^2 \,,
    \label{sm:isoperimetric}
\end{equation}
where we use the convention $n + 1 \equiv 1$. Equality holds if and only if the points $(a_1, b_1), \allowbreak \dots,\allowbreak (a_n, b_n)$ form a regular $n$-sided polygon of any size on the $a$--$b$ plane in this order.

\paragraph{Proof}
Let $s_i $ indicate the line segment connecting $(a_i , b_i )$ and $(a_{i+1}, b_{i+1})$ in the $a$--$b$ plane. Consider the polygon $P$ formed by the sequence of edges $(s_1,\dots,s_n)$, which may be non-simple, i.e., self-intersecting. The length of the perimeter of $P$ is given by $L(P) \coloneqq \sum_i L_{i+1,i}  $, and the signed area of a polygon $P$ is defined as $\Omega (P) \coloneqq \sum_i \Omega_{i+1,i} $ \citep{radoIsoperimetricInequalityLebesgue1947}.
The signed area has been shown to be equivalent to \citep[section 6.4,][]{radoLemmaTopologicalIndex1936}
\begin{equation}
    \Omega(P)=\iint_{\mathbb R^2} w_{P}(a,b)\,da\,db,
    \label{sm:winding_number}
\end{equation}
where $w_{P}(a,b)$ is the winding number of the curve $P$ at $(a,b)$, i.e., the number of times the curve $P$ circulates around the point $(a,b)$ in the clockwise direction subtracted from that in the counter-clockwise direction. 

We now define another polygon $Q$, which is called the ``convexification'' of $P$ \citep{scottFamilyPolygons1982,faryIsoperimetryVariableMetric1982}. $Q$ is constructed from the same set of line segments as edges, $(s_{1},\dots,s_n)$, but the edges are translated and permuted so that their angle monotonically increases in a counter-clockwise fashion. Formally, $Q$ is formed by translating and connecting the sequence of edges $(s_{\sigma(1)},\dots,s_{\sigma(n)})$, where $\sigma$ is a permutation that guarantees that the angle of the line segment $(a_{\sigma(i)}, b_{\sigma(i)})$--$(a_{\sigma(i+1)}, b_{\sigma(i+1)})$ relative to the $a$ axis is monotonically increasing in $i$.

Since the edges of $Q$ can only increase in angle, $Q$ is convex
and simple. For a simple polygon with its edges arranged
in the counter-clockwise direction such as $Q$, $\Omega(Q)$
is the regular geometric area, which obeys the well-known isoperimetric
inequality \citep{fanAlgebraicProofIsoperimetric1955},
\begin{equation}
    4 n \tan(\pi/n) \, \Omega(Q)
    \leq L(Q)^{2}.
    \label{sm:iso_Q}
\end{equation}
Since $Q$ and $P$ have the same number of edges of the same length, $L(Q)=L(P)$. As for the signed area, it has been proved that $\Omega(P) \leq \Omega(Q)$ for the expression of the signed area in Eq.~\eqref{sm:winding_number} \citep[Lemma 1,][]{faryIsoperimetryVariableMetric1982}. Plugging these relations into Eq.~\eqref{sm:iso_Q} gives 
\begin{equation}
    4n \tan(\pi/n) \, \Omega (P)
    \leq L(P)^{2}.
    \label{sm:iso_proof1}
\end{equation}

Next, let $P'$ be the polygon obtained by interchanging $a$ and $b$ from $P$, which has the signed area $\Omega(P') = -\Omega(P)$ and the perimeter $L(P') = L(P)$. Therefore, Eq.~\eqref{sm:iso_proof1} applied to $P'$ gives
\begin{equation}
    - 4n \tan(\pi/n) \, \Omega (P)
    \leq L(P)^{2}.
    \label{sm:iso_proof2}
\end{equation}
Combining Eqs.~\eqref{sm:iso_proof1} and \eqref{sm:iso_proof2} gives the desired result~\eqref{sm:isoperimetric}.

The isoperimetric inequality for $Q$~\eqref{sm:iso_Q} is saturated if and only if $Q$ is regular~\cite{fanAlgebraicProofIsoperimetric1955}. The inequality $\Omega(P) \leq \Omega(Q)$ is saturated if and only if $P$ is convex and the edges are counter-clockwise oriented, or $P$ is contained in a straight line~\cite{faryIsoperimetryVariableMetric1982}. Therefore, the equality condition of Eq.~\eqref{sm:iso_proof1} is that $P$ is regular, and its edges are counter-clockwise oriented.
Similarly, the equality condition of Eq.~\eqref{sm:iso_proof2} is that $P$ is regular, and its edges are clockwise oriented. The equality in Eq.~\eqref{sm:isoperimetric} holds for both of these cases.
\qed 

We note that the statement is purely algebraic, although the proof relies on geometric concepts. A direct consequence of this theorem is the following corollary, which may give a tighter bound.

\paragraph{Corollary}
For a sequence of $n$ points $(a_1, b_1), \allowbreak \dots,\allowbreak (a_n, b_n)$ on $\mathbb R^2$, let $n'$ be the number of times the joint value $(a,b)$ changes over the cyclic sequence. Formally, $n'$ is the number of labels $i\in\{1,\dots, n\}$ such that $(a_i, b_i) \neq (a_{i+1}, b_{i+1})$, where we use the convention $n+1 \equiv 1$. Then,
\begin{equation}
    \qty(4n' \tan \frac{\pi}{n'} )\,
    \Bvert{ \sum_{i=1}^n \Omega_{i+1,i} }
    \leq \Bparen{ \sum_{i=1}^n  L_{i+1,i} } ^2 \,.
    \label{sm:isoperimetric_tighter}
\end{equation}

\paragraph{Proof}
This corollary is essentially because the $n$ points form an $n'$-sided polygon. To prove it formally, Let $I = \Set{ i\in\{1,\dots,n \} | (a_i, b_i) \neq (a_{i+1}, b_{i+1})}$ be the partial set of labels. The size of this set is $\vert I\vert=n'$. The isoperimetric inequality~\eqref{sm:isoperimetric} applied to the sequence of points $\Set{ (a_i,b_i) | i\in I}$ (in the increasing order in $i$) gives
\begin{equation}
    \qty(4n' \tan \frac{\pi}{n'} )\,
    \Bvert{ \sum_{i\in I} \Omega_{i+1,i} }
    \leq \Bparen{ \sum_{i\in I}  L_{i+1,i} } ^2.
    \label{sm:isoperimetric_tighter_proof1}
\end{equation}
On the other hand, since $\Omega_{i+1,i}=0$ and $L_{i+1,i}=0$ for $i\notin I$, 
\begin{equation}
    \sum_{i=1}^n \Omega_{i+1,i} = \sum_{i\in I} \Omega_{i+1,i},\quad 
    \sum_{i=1}^n L_{i+1,i} = \sum_{i\in I} L_{i+1,i}.
    \label{sm:isoperimetric_tighter_proof2}
\end{equation}
Combining Eqs.~\eqref{sm:isoperimetric_tighter_proof1} and \eqref{sm:isoperimetric_tighter_proof2} proves the desired result.
\phantom{A}\qed

We introduce another Lemma, which will be used for bipartite cases. 

\paragraph{Lemma}
Consider a sequence of $n$ points $(a_1, b_1), \allowbreak \dots,\allowbreak (a_n, b_n)$ on $\mathbb R^2$ that satisfies $(a_i - a_{i+1})(b_i - b_{i+1}) = 0$ for $i=1,\dots,n$. In other words, the value of $a$ and $b$ does not simultaneously change along the sequence. Then, 
\begin{equation}
    16 \,
    \Bvert{  \sum_{i=1}^n \Omega_{i+1,i} }
    \leq \Bparen{ \sum_{i=1}^n  L_{i+1,i} }^2\,.
    \label{sm:isoperimetric_bipartite}
\end{equation}

\paragraph{Proof}
The proof proceeds similarly to the proof of Eq.~\eqref{sm:isoperimetric}. We define the edges $s_i$ and the polygon $P$ as in the previous proof. Due to the assumption, the direction of each edge is either $0$, $\pi/2$, $\pi$, or $3\pi/2$ relative to the $a$ axis.

Let $Q$ be the convexification of $P$. Then, $Q$ must be a rectangle, and hence satisfy the inequality~\eqref{sm:iso_Q} with $n=4$:
\begin{equation}
    16 \,\Omega(Q)
    \leq L(Q)^{2}.
\end{equation}
As in the previous proof, we have $L(P) = L(Q)$ and $\Omega(P) \leq \Omega(Q)$, and therefore 
\begin{equation}
    16 \,\Omega(P)
    \leq L(P)^{2}
\end{equation}
holds for $P$. The proof for $-\Omega(P)$ proceeds similarly to the previous proof.
\qed

\subsection{Proof of the main results}
\label{subsec:final}

Finally, we combine these tools to finish the proof of Eqs.~\eqref{eq:main}, \eqref{eq:main_tighter}, and \eqref{eq:main_bipartite}. Here we explicitly prove the tighter version Eq.~\eqref{eq:main_tighter}, which applies to the general setup, and the bipartite case Eq.~\eqref{eq:main_bipartite}. The simpler version Eq.~\eqref{eq:main} follows from Eq.~\eqref{eq:main_tighter} by using $n'_c \tan(\pi/n'_c) \geq n_c \tan(\pi/n_c)$ for $n'_c \leq n_c$ and that maximizing over all simple cycles gives a larger result than maximizing over the restricted set $\CS$.

\paragraph{Proof of Eq.~\eqref{eq:main_tighter}}
First, we apply the corollary~\eqref{sm:isoperimetric_tighter} of the isoperimetric inequality to the sequence of points $(a_{i_1}, b_{i_1}), \dots, (a_{i_{n_c}}, b_{i_{n_c}})$ for each cycle $c=(i_1 \,{\to}\, \cdots \,{\to}\, i_{n_c} \,{\to}\, i_1 ) $. This gives
\begin{equation}
    \qty(4n'_c \tan \frac{\pi}{n'_c} )\,
    \Bvert{ \sum_{e\in c} \Omega_e }
    \leq \Bparen{ \sum_{e\in c}  L_e } ^2.
    \label{sm:final_isoperimetric}
\end{equation}
We combine Eq.~\eqref{sm:final_isoperimetric} with Eq.~\eqref{sm:numerator} to obtain
\begin{align}
    \Bvert{\sum_{i>j} \cur_{ij}\Omega_{ij}}
    &\leq \sum_{c\in \CS} \cur_{c} \, \Bvert{\sum_{e\in c} \Omega_{e} }
    \notag \\
    &\leq \sum_{c\in \CS} \cur_c \, \Bparen{ \sum_{e\in c} L_e }^2 \qty( 4n'_c \tan\frac{\pi}{n'_c} ) ^{-1},
    \label{sm:final_numerator}
\end{align}
where we used that $\cur_{c} > 0$. Next, we combine Eq.~\eqref{sm:denomenator} with the generalized TUR~\eqref{sm:genTUR_multicyclic} applied to $X_e = L_e$ to rewrite
\begin{align}
    \sum_{i>j}\act_{ij}L^2_{ij}
    &\geq
    \sum_{c\in \CS} \cur_{c} \, \Bparen{ \sum_{e\in c}  \frac{\act_e } { \cur_e } L^2_e }
    \notag \\
    &\geq
    \sum_{c\in \CS} \cur_c \, \Bparen{ \sum_{e\in c} L_e }^2 \qty ( n_c \tanh \frac{\aff_c}{2n_c} )^{-1}.
    \label{sm:final_denominator}
\end{align}
Plugging Eqs.~\eqref{sm:final_numerator} and \eqref{sm:final_denominator} into the expression of $\chi_{ba}$ in Eq.~\eqref{sm:chi_convenient} gives
\begin{equation}
    \vert \chi_{ba} \vert
    \leq \frac{4 \sum_{c\in\CS}   \cur_c (\sum_{e\in c}  L_e)^2 [4n'_c \tan(\pi / n'_c)]^{-1}}
    {\sum_{c\in\CS}  \cur_c (\sum_{e\in c} L_e)^2 [n_c \tanh(\aff_c/2n_c)]^{-1}} .
\end{equation}
Finally, we use the inequality
\begin{equation}
    \frac{\sum_{c\in \CS} y_{c}} {\sum_{c\in \CS} x_{c}}\leq
    \frac {\sum_{c\in \CS}x_c\max_{c\in \CS} \frac{y_{c}}{x_{c}} }{\sum_{c\in \CS} x_{c}}
    =\max_{c\in \CS} \frac{y_{c}}{x_{c}}
    \label{sm:inequality_max}
\end{equation}
for any $x_{c}\in\mathbb{R}_{>0}$ and $y_{c}\in\mathbb{R}$, which is saturated if and only if $y_c / x_c$ are equal across all $ c \in \CS$. This leads to
\begin{align}
    \vert \chi_{ba} \vert
    &\leq \max_{c\in \CS} \frac{4  \cur_c (\sum_{e\in c}  L_e)^2 [4n'_c \tan(\pi / n'_c)]^{-1}}
    { \cur_c (\sum_{e\in c} L_e)^2 [n_c \tanh(\aff_c/2n_c)]^{-1}} 
    \notag \\
    &= \max_{c\in \CS} \frac{n_c\tanh(\aff_c/2n_c)}{n'_c\tan(\pi/n'_c)},
    \label{sm:final_proof}
\end{align}
which proves the desired inequality in Eq.~\eqref{eq:main_tighter}.
\qed

\paragraph{Proof of Eq.~\eqref{eq:main_bipartite}}
Similarly to the proof of Eq.~\eqref{eq:main_tighter}, we apply the isoperimetric inequality for bipartite observables, Eq.~\eqref{sm:isoperimetric_bipartite}, to the sequence of points $(a_{i_1}, b_{i_1}), \dots, (a_{i_{n_c}}, b_{i_{n_c}})$ for each cycle $c=(i_1 \,{\to}\, \cdots \,{\to}\, i_{n_c} \,{\to}\, i_1 ) $. This gives 
\begin{equation}
    16\, 
    \Bvert{ \sum_{e\in c} \Omega_e }
    \leq \Bparen{ \sum_{e\in c}  L_e } ^2.
\end{equation}
This is formally equivalent to Eq.~\eqref{sm:final_isoperimetric} with the replacement of $n'_c$ with 4. Therefore, with the same lines of reasoning as in the previous proof, we obtain Eq.~\eqref{sm:final_proof} with the replacement of $n'_c$ with 4. This is identical to the desired result~\eqref{eq:main_bipartite}. 
\phantom{A}\qed

\section{Bound for finite time lag \texorpdfstring{$\tau$}{tau}}
\label{sec:finitetime}

While our main result concerns the correlations in the short-$\tau$ region, here we conjecture that a similar result holds for any finite $\tau$ based on numerical evidence. More precisely, we introduce a finite-$\tau$ version of Eq.~\eqref{eq:asymm_def} as
\begin{equation}
    \chi_{ba}^{\tau} = \frac{C_{ba}^{\tau}-C_{ab}^{\tau}}{2\sqrt{(\Delta_{\tau}C_{aa})(\Delta_{\tau}C_{bb})}}.
\end{equation}
For this ratio, we conjecture
\begin{equation}
    \vert\chi_{ba}^{\tau}\vert\leq\max_{c}\frac{\tanh(\aff_{c}/2n_{c})}{\tan(\pi/n_{c})}\le\max_{c}\frac{\aff_{c}}{2\pi}
    \label{sm:finite}
\end{equation}
for all $\tau>0$. This conjecture means that the maximum cycle affinity
limits the asymmetry of the cross-correlation $C_{ba}^{\tau}-C_{ab}^{\tau}$
for all time lags, not just the short-$\tau$ regime. Equation~\eqref{sm:finite} implies the main result~\eqref{eq:main} by taking the limit $\lim_{\tau\to0} \chi_{ba}^{\tau} = \chi_{ba}$. 

\begin{figure}[t]
    \centering
    \includegraphics[width=\hsize,clip]{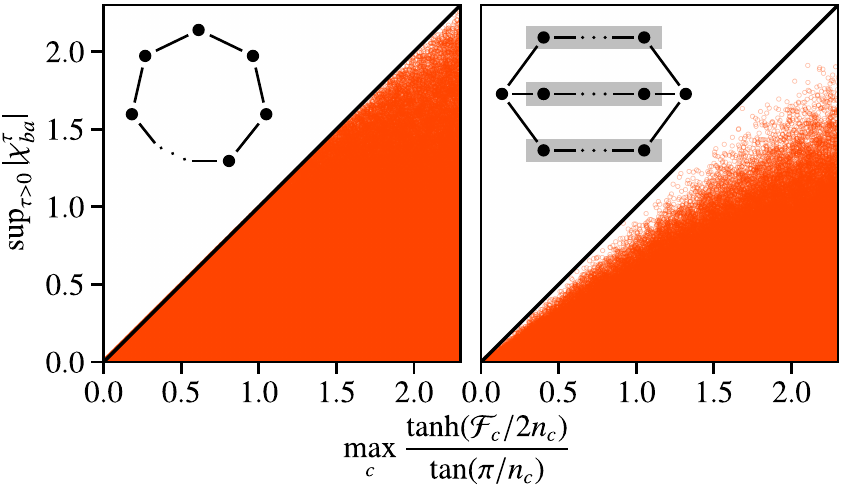}
    \caption{Numerical evidence for the finite-$\tau$ bound~\eqref{sm:finite_sup} for unicyclic systems (left) and multicyclic systems (right). We generate $10^{6}$ systems with random numbers of states, random transition rates, and random observables $a$ and $b$, and we plot both sides of Eq.~\eqref{sm:finite_sup}. Insets show the topology of the systems. A circle $\bullet$ denotes a state, and a line \textbf{---} denotes an edge with a nonzero transition rate. The number of states in unicyclic systems (left) are chosen randomly from 3 to 16, and the number of states in each gray-shaded area in multicyclic systems (right) is chosen randomly from 1 to 8. }
    \label{fig:finitetime}
\end{figure}

To numerically test Eq.~\eqref{sm:finite}, we note that Eq.~\eqref{sm:finite} for all $\tau > 0 $ is equivalent to 
\begin{equation}
    \sup_{\tau>0}\vert\chi_{ba}^{\tau}\vert\leq\max_{c}\frac{\tanh(\aff_{c}/2n_{c})}{\tan(\pi/n_{c})}.
    \label{sm:finite_sup}
\end{equation}
We plot both sides of Eq.~\eqref{sm:finite_sup} for $10^{6}$ randomly generated systems in Fig.~\ref{fig:finitetime}. All the points obey the bound Eq.~\eqref{sm:finite_sup} without a single exception, which provides numerical evidence to the bound~\eqref{sm:finite_sup}. We leave analytical investigation of this finite-$\tau$ bound to future work.

\section{Supplemental discussions}
\label{sec:additional}

\subsection{Implicit assumptions}

We clarify two assumptions that are implicitly made so that the main results are well-defined.
First, the denominator of $\chi_{ba}$ should be nonzero, i.e., $\sum_{i,j} \flux_{ij} a_i a_j \neq 0 $ and $\sum_{i,j} \flux_{ij} b_i b_j \neq 0$, when we define $\chi_{ba}$. Second, the system should have at least one cycle because otherwise the right-hand side of the result is ill-defined.

These assumptions are not very restrictive because they are automatically satisfied whenever the cross-correlations are asymmetric, $C^\tau_{ba} \neq C^\tau_{ab}$. In other words, if one of these assumptions are violated, it follows that $C^\tau_{ba} = C^\tau_{ab}$.
Indeed, if $\sum_{i,j} \flux_{ij} a_i a_j = 0 $ holds, Eq.~\eqref{eq:dCaaDef} shows that $a_i = a_j$ for any pair of $i, j$ with nonzero flux $\flux_{ij}\neq 0$. 
This is equivalent to $a$ being constant along any steady-state trajectory, which leads to $C^\tau_{ba}=C^0_{ba}$. Therefore, the asymmetry is $C^\tau_{ba} - C^\tau_{ab} = C^0_{ba} - C^0_{ab} = 0$.
Similarly, if $\sum_{i,j} \flux_{ij} b_i b_j = 0 $ holds, the asymmetry is zero.
For a system with no cycles, its steady state is always detailed-balanced, $\flux_{ij} = \flux_{ji}$~\cite{schnakenberg1976network}. Therefore, the system has no asymmetry, $C^\tau_{ba} - C^\tau_{ab} = 0 $, as confirmed from Eq.~\eqref{eq:CbaDef}.

\subsection{Rescaling observables}
\label{subsec:rescale}

In Sec.~\ref{sec:derivation}, we have proven our result based on the   assumption~\eqref{sm:rescale_assumption}. Imposing this assumption does not lose generality, as discussed in Sec.~\ref{subsec:rewrite}, and therefore the proof is already completed. Nevertheless, it will be helpful to give an explicit  proof of Eq.~\eqref{eq:main_tighter} for general pair of observables $a$ and $b$.

Let $a$ and $b$ be any pair of observables that does not necessarily satisfy the assumption~\eqref{sm:rescale_assumption}. We define a constant
\begin{equation}
    \delta = \sqrt{\frac{ \sum_{i,j} \flux_{ij} b_i b_j}{ \sum_{i,j} \flux_{ij} a_i a_j}},
    \label{sm:coefficient}
\end{equation}
and introduce a new pair of observables $a'_i = \delta a_i$ and $b'_i = b_i$. One can easily confirm the condition $ \sum_{i,j} \flux_{ij} a'_i a'_j = \sum_{i,j} \flux_{ij} b'_i b'_j$, and therefore the above proof applies to the pair $a'$ and $b'$ to deduce
\begin{equation}
    \vert \chi_{b'a'} \vert
    \leq \max_{c\in \CS} \frac{n_c\tanh(\aff_c/2n_c)}{n'_c\tan(\pi/n'_c)}.
    \label{sm:rescale_prime}
\end{equation}
Since $C^\tau_{b'a'} = \delta C^\tau_{ba}$, $C^\tau_{a'b'} = \delta C^\tau_{ab}$, $C^\tau_{a'a'} = \delta^2 C^\tau_{aa}$, and $C^\tau_{b'b'} = C^\tau_{bb}$, it follows that $\vert\chi_{ba}\vert = \vert\chi_{b'a'}\vert$. Combining this with Eq.~\eqref{sm:rescale_prime} completes the explicit proof of the main result~\eqref{eq:main_tighter}.

\subsection{Equality conditions}

We discuss the equality condition of our main result, Eq.~\eqref{eq:main}, for unicyclic systems. 

\paragraph{Proposition}
For unicyclic systems, the first (tighter) inequality in our main result~\eqref{eq:main} is saturated if and only if (i) $\act_e$ is uniform across the cycle, and (ii) there exist a scaling constant $\gamma$ such that the points $(\gamma a_1, b_1), \allowbreak \dots, (\gamma a_n, b_n)$ form a regular $n$-sided polygon in this order.

\paragraph{Proof} 
We first find the equality condition under the assumption~\eqref{sm:rescale_assumption}, and then recast it to the general case. 

For $a$ and $b$ obeying the  assumption~\eqref{sm:rescale_assumption}, our main result~\eqref{eq:main} for unicyclic systems is simply the combination of the generalized TUR and the isoperimetric inequality, as discussed in the main text. By combining their equality conditions, which are stated below Eq.~\eqref{sm:genTUR_multicyclic} and Eq.~\eqref{sm:isoperimetric}, and noting that the current $\cur_e$ is always uniform for unicyclic systems, one can easily find that the equality of Eq.~\eqref{eq:main} holds if and only if $\act_e$ is uniform across the cycle, and $(a_1, b_1),\dots, (a_n, b_n)$ form a regular $n$-sided polygon in this order.

For $a$ and $b$ that do not necessarily obey the assumption~\eqref{sm:rescale_assumption}, the result~\eqref{eq:main} is obtained by applying the above proof to the rescaled variables $a' = \delta a$ and $b' = b$, as discussed in Sec.~\ref{subsec:rescale}. Therefore, the equality condition is that (i) $\act_e$ is uniform along the cycle, and (ii') $(\delta a_1 , b_1),\dots ,(\delta a_n , b_n)$ form a regular $n$-sided polygon in this order. It remains to show that the combination of (i) and (ii') is equivalent to the combination of (i) and (ii) in the statement. We can assume without loss of generality that $\gamma > 0$ because, whenever $(\gamma a_1, b_1), \allowbreak \dots, (\gamma a_n, b_n)$ form a regular $n$-sided polygon, so do $(-\gamma a_1, b_1), \allowbreak \dots, (-\gamma a_n, b_n)$. 

When (i) and (ii') holds, (ii) immediately holds with the choice $\gamma = \delta$. Conversely, suppose that (i) and (ii) are satisfied. Then, the coefficient $\delta$ in Eq.~\eqref{sm:coefficient} is calculated as
\begin{align}
    \delta &= \sqrt{ \frac{ \sum_{i} \act_{i+1,i} (b_i - b_{i+1})^2 }{\sum_{i} \act_{i+1,i} (a_i - a_{i+1})^2 }  }
    \notag \\*
    &= \gamma \sqrt{ \frac{\sum_{i} (b_i - b_{i+1})^2 }{ \sum_{i} (\gamma a_i - \gamma a_{i+1})^2 }  } 
    \notag \\*
    &= \gamma,
    \label{sm:delta_gamma}
\end{align}
where the first equality is due to Eq.~\eqref{eq:dCaaDef}, and the second equality is due to the condition (i). The last equality is confirmed by expressing $\gamma a_i = r \cos( i\cdot 2\pi/n + \theta_0 )$ and $b_i = r \sin( i\cdot 2\pi/n + \theta_0)$ for any $r$ and $\theta_0$ and calculating
\begin{equation}
    \sum_i (b_i - b_{i+1})^2 
    = \sum_i (\gamma a_i - \gamma a_{i+1})^2 
    = 2nr^2  \sin^2 \frac{\pi}{n}
\end{equation}
using the sum-to-product identities and the power-reduction formulae from trigonometry. 
Equation~\eqref{sm:delta_gamma} combined with (ii) implies (ii').
\qed

Note that the condition (i) is equivalent to the uniformity of the one-way fluxes $\flux_{21} = \flux_{32} = \dots = \flux_{1n}$ and $\flux_{12} = \flux_{23} = \dots = \flux_{n1}$. This is implied by, but not equivalent to, the uniformity of the rates $R_{21} = R_{32} = \dots = R_{1n}$ and $R_{12} = R_{23} = \dots = R_{n1}$. 

We can similarly discuss the equality condition for multicyclic systems. For observables that satisfy the assumption~\eqref{sm:rescale_assumption}, the equality condition is obtained by considering the equality conditions of Eq.~\eqref{sm:numerator}, Eq.~\eqref{sm:denomenator}, the generalized TUR~\eqref{sm:genTUR_multicyclic}, the isoperimetric inequality~\eqref{sm:isoperimetric}, and Eq.~\eqref{sm:inequality_max}. For general observables, we can obtain the equality condition via a rescaling argument as above. We omit further details. 


\end{document}